\documentclass[reprint,floatfix,amsmath,amssymb,aps,prb,superscriptaddress]{revtex4-2}

\usepackage{graphicx}
\usepackage{dcolumn}
\usepackage{bm}

\preprint{APS/123-QED}

\usepackage[separate-uncertainty=true]{siunitx}
\usepackage[version=4]{mhchem}
\usepackage{float}
\usepackage{mathtools}
\usepackage{hyperref}
\DeclareSIUnit\samples{\text{S}}
\usepackage{nicematrix}
\usepackage{nicefrac}
\usepackage{booktabs}
\newcommand{\ra}[1]{\renewcommand{\arraystretch}{#1}}

\usepackage{silence}
\begin{document}
\title{Extracting Photon-Number Information from Superconducting Nanowire Single-Photon Detectors Traces via Mean-Derivative Projection}

\author{I.S. Kuijf}
\thanks{These authors contributed equally.}
\affiliation{Lorentz Institute and Leiden Institute of Advanced Computer Science, Leiden University, P.O. Box 9506, 2300 RA Leiden, The Netherlands}

\author{F.B. Baalbergen}
\thanks{These authors contributed equally.}
\affiliation{Huygens-Kamerlingh Onnes Laboratory, Leiden University, P.O. Box 9504, 2300 RA Leiden, The Netherlands}

\author{L. Seldenthuis}
\thanks{These authors contributed equally.}
\affiliation{Huygens-Kamerlingh Onnes Laboratory, Leiden University, P.O. Box 9504, 2300 RA Leiden, The Netherlands}

\author{E.P.L. van Nieuwenburg}
\affiliation{Lorentz Institute and Leiden Institute of Advanced Computer Science, Leiden University, P.O. Box 9506, 2300 RA Leiden, The Netherlands}

\author{M.J.A. de Dood}
\affiliation{Huygens-Kamerlingh Onnes Laboratory, Leiden University, P.O. Box 9504, 2300 RA Leiden, The Netherlands}

\date{\today}

\begin{abstract}
Photon-number resolved detection with superconducting nanowire single-photon detectors (SNSPDs) attracts increasing interest, but lacks a systematic framework for interpreting and benchmarking this capability.
In this work, we combine principal component analysis (PCA) with a new readout technique to explore the photon-number resolving capabilities of SNSPDs and find that the information of the photon number is contained in a single principal component which approximates the time derivative of the average response trace.
We introduce a new confidence metric based on the Bhattacharyya coefficient to quantify the photon-number-resolving capabilities of a detector system and show that this metric can be used to compare different systems.
Our analysis and interpretation of the principal components imply that photon-number resolution in SNSPDs can be achieved with moderate hardware requirements in terms of both sample rate (5 GSample/sec) and analog bandwidth (3 GHz) and could be implemented in an FPGA, giving a highly scalable solution for real-time photon counting.
\end{abstract}

\maketitle

\section{Introduction}
Photon-number-resolving detectors are critical components for photonic quantum technologies.
The ability to detect and distinguish between different numbers of photons with high fidelity underpins a wide range of quantum information protocols and is deemed essential for applications such as linear optical quantum computing~\cite{You2020, Gruenenfelder2023}, quantum metrology~\cite{Bhargav2021}, and boson sampling~\cite{Zhong2020}.
Both high detection efficiency and accurate timing information are important for detecting quantum states of light with high fidelity and maintaining scalability in multi-photon experiments.
Superconducting nanowire single-photon detectors (SNSPDs) act as single-photon threshold detectors that satisfy both requirements with system detection efficiencies exceeding 90\%~\cite{Chang2021, Reddy2020, Hu2020} combined with reset times of \qtyrange{10}{50}{\nano\second}~\cite{Kerman2013, Zadeh2017, EsmaeilZadeh2020} and timing jitter below \qty{20}{\pico\second} in commercially available systems~\cite{Los2024, EsmaeilZadeh2020, Chang2019}.

Single-photon threshold detectors, such as single-photon avalanche photodiodes (SPADs) and conventional SNSPDs, traditionally only register the presence of at least one photon and do not count the number of photons.
While this can be circumvented by using multiplexed detector arrays~\cite{Stasi2023, Stasi2024}, an order of $n^2$ detectors are required to detect a state with $n$ photons~\cite{Jnsson2019}.
Recent work shows that SNSPDs themselves exhibit intrinsic photon number resolving capabilities~\cite{Cahall2017, Zhu2020, Schapeler2024,Sempere2022, Los2024}, removing the need for multiplexing schemes.
Cahall et al.~\cite{Cahall2017} were the first to report photon-number dependence in SNSPDs, by measuring variations in the rise time of the SNSPD output signal that were attributed to differences in the initial hotspot resistance, which in turn depends on the photon number.
This work was followed by several others exploring different strategies for extracting photon-number information from the analog signal, including linear fits to the slope of the rising edge ~\cite{Sempere2022}, amplitude correlations~\cite{Zhu2020}, and measurements of the arrival time relative to a trigger signal~\cite{Los2024}.

Recently, Schapeler et al.~\cite{Schapeler2024} demonstrated that most of the photon-number-resolving information is concentrated within the rising edge of the electrical signal and can be extracted using a time-to-digital converter conditioned on a threshold.
In their work, they compared measurements of the rising and falling edge arrival times with a principal component analysis (PCA) of the full electrical output signal of the detector.
PCA is a technique widely used in multivariate statistics~\cite{abdi2010principal}, that allows for identifying the most informative directions in the high-dimensional signal space, effectively reducing the complexity of the problem while preserving the photon-number-discriminating features.

Although this recent application of PCA successfully identifies the key discriminating regions of the signal trace, the physical interpretation of the principal components remains unknown.
Understanding the contributions of different parts of the response trace is essential for optimizing the photon-number resolving capability of SNSPDs.
Furthermore, it remains an open question what the best method is to quantify and compare the photon-number resolving capability across different detectors or configurations with different proposed methods~\cite{Humphreys2015, schapeler2025_2}.
This gap in understanding limits our ability to define the minimally sufficient requirements needed for the readout hardware to do photon-number resolved detection with SNSPDs.

In this work, we address these questions by applying PCA to the electrical response traces of a commercially available SNSPD system.
We show that the first principal component is equivalent to the derivative of the mean response trace.
We present a physically intuitive interpretation of the PCA results and demonstrate how it can be used to optimally distinguish between different photon-numbers.
Our method extracts photon-number information by utilizing the entire measured response trace, as opposed to only using one or two key points of the trace.
We show that this optimal photon number determination can be done in an efficient and scalable way with modest hardware requirements, enabling real-time photon number resolution.
Furthermore, we introduce a confidence metric that quantifies the photon-number-resolving capability of a detector system.
This metric provides a new standard for evaluating and comparing detectors in terms of their suitability for photon-number resolving applications.

\section{Data acquisition}

\begin{figure}
\centering
\begingroup
\makeatletter
\providecommand\color[2][]{
\errmessage{(Inkscape) Color is used for the text in Inkscape, but the package 'color.sty' is not loaded}
\renewcommand\color[2][]{}
}
\providecommand\transparent[1]{
\errmessage{(Inkscape) Transparency is used (non-zero) for the text in Inkscape, but the package 'transparent.sty' is not loaded}
\renewcommand\transparent[1]{}
}
\providecommand\rotatebox[2]{#2}
\newcommand*\fsize{\dimexpr\f@size pt\relax}
\newcommand*\lineheight[1]{\fontsize{\fsize}{#1\fsize}\selectfont}
\ifx\svgwidth\undefined
\setlength{\unitlength}{235.85674136bp}
\ifx\svgscale\undefined
\relax
\else
\setlength{\unitlength}{\unitlength * \real{\svgscale}}
\fi
\else
\setlength{\unitlength}{\svgwidth}
\fi
\global\let\svgwidth\undefined
\global\let\svgscale\undefined
\makeatother
\begin{picture}(1,0.65825184)
\lineheight{1}
\setlength\tabcolsep{0pt}
\put(0,0){\includegraphics[width=\unitlength,page=1]{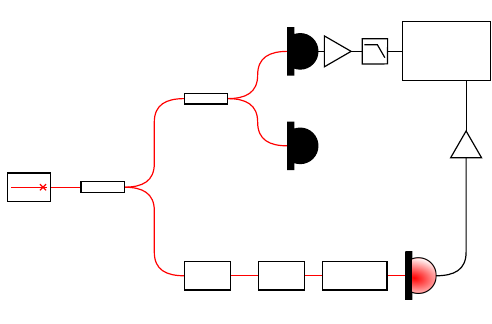}}
\put(0.06308894,0.19793037){\color[rgb]{0,0,0}\makebox(0,0)[t]{\smash{\begin{tabular}[t]{c}Pulsed\\laser\end{tabular}}}}
\put(0.60923419,0.62238264){\color[rgb]{0,0,0}\makebox(0,0)[t]{\smash{\begin{tabular}[t]{c}Photodiode\end{tabular}}}}
\put(0.90804887,0.54171441){\color[rgb]{0,0,0}\makebox(0,0)[t]{\smash{\begin{tabular}[t]{c}Digitizer\end{tabular}}}}
\put(0.20893657,0.30012389){\color[rgb]{0,0,0}\makebox(0,0)[t]{\smash{\begin{tabular}[t]{c}$1:1000$\end{tabular}}}}
\put(0.41926038,0.47704582){\color[rgb]{0,0,0}\makebox(0,0)[t]{\smash{\begin{tabular}[t]{c}$1:1$\end{tabular}}}}
\put(0.37335325,0.35710536){\color[rgb]{0,0,0}\makebox(0,0)[t]{\smash{\begin{tabular}[t]{c}$1000$\end{tabular}}}}
\put(0.33999628,0.17523163){\color[rgb]{0,0,0}\makebox(0,0)[t]{\smash{\begin{tabular}[t]{c}$1$\end{tabular}}}}
\put(0.42212302,0.08441439){\color[rgb]{0,0,0}\makebox(0,0)[t]{\smash{\begin{tabular}[t]{c}VOA\end{tabular}}}}
\put(0.57235435,0.08441439){\color[rgb]{0,0,0}\makebox(0,0)[t]{\smash{\begin{tabular}[t]{c}VOA\end{tabular}}}}
\put(0.72249517,0.08646228){\color[rgb]{0,0,0}\makebox(0,0)[t]{\smash{\begin{tabular}[t]{c}$-20$dB\end{tabular}}}}
\put(0.83684625,0.00086495){\color[rgb]{0,0,0}\makebox(0,0)[t]{\smash{\begin{tabular}[t]{c}SNSPD\end{tabular}}}}
\put(0.81280924,0.37343864){\color[rgb]{0,0,0}\makebox(0,0)[t]{\smash{\begin{tabular}[t]{c}Cryogenic\\amplifier\end{tabular}}}}
\put(0.59647618,0.26527208){\color[rgb]{0,0,0}\makebox(0,0)[t]{\smash{\begin{tabular}[t]{c}Power meter\end{tabular}}}}
\end{picture}
\endgroup

\caption{Schematic overview of the experimental setup showing the laser, attenuators and readout electronics.}
\label{fig:setup}
\end{figure}

Detection events are collected using a commercial fiber-coupled SNSPD system (ID~Quantique ID281), operated at \qty{2}{\kelvin} in a closed-cycle refrigerator with cryogenic amplification at \qty{40}{\kelvin} with unknown amplification and bandwidth.
This detector is designed for a wavelength of \qty{1550}{\nm} and is biased at the manufacturer-recommended setting.
A schematic overview of the measurement setup is shown in \autoref{fig:setup}.
To generate events, the detector is illuminated with attenuated laser pulses from a fiber-coupled laser diode (ID~Quantique ID 3000) with a central wavelength of \qty{1551}{\nm} and a pulse length of \qty{23}{\pico\second}.
The laser repetition rate is set to \qty{500}{\kilo\hertz} to ensure operation in a regime where the time between pulses is much larger than the thermal and electronic reset times of the SNSPD, thus not compromising the detection efficiency.

Light from the laser is coupled to a 1000:1 beam splitter, where the high-power output is divided equally between a fast photodiode for synchronization and a power meter (Thorlabs S122C) for monitoring the laser power.
The low-power output is connected to two fiber-coupled variable optical attenuators (Thorlabs V1550F) and a fixed \qty{20}{\dB} attenuator (Thorlabs FA20T), before the light is coupled to the SNSPD.
The variable optical attenuators are controlled using a voltage set by a computer, using the outputs of a digital-to-analog converter (National Instruments NI USB-6341).
The transmission of the total system is carefully calibrated beforehand so that the set voltages on the variable optical attenuators and the measured power on the power meter are sufficient to calculate the mean photon number per pulse on the SNSPD.

The voltage signals from both the photodiode and SNSPD are sampled synchronously using two channels of a digitizer (Teledyne ADQ7DC) with an analog bandwidth of \qty{3}{\giga\hertz} and a sample rate of \qty{5}{\giga\samples\per\second}.
The analog bandwidth is defined as the frequency where the signal is attenuated by \qty{3}{\deci\bel}.
Aliasing is not of concern due to the expected frequency response of SNSPDs consisting of mostly frequencies below \qty{2}{\giga\hertz}~\cite{Kerman2006, Zhu2019}.
The signal from the photodiode is used to trigger the digitizer, and is filtered with a \qty{190}{\mega\hertz} low-pass filter before being digitized.
This stretches the pulse in the time-domain and removes high frequency noise from the signal to allow for a more precise determination of the arrival time of the laser pulse in post-processing by fitting the peak position over more datapoints.
The electronic SNSPD signal is attenuated with a \qty{3}{\dB} attenuator before sending it to the digitizer, to prevent signal clipping.

To form the dataset, \num{100000} light traces of the SNSPD and the corresponding photodiode signals are recorded at 20 different mean photon numbers, resulting in a total of \num{2000000} measurements.
The mean photon numbers range from $\mu=0.003$ to $\mu=2.55$ photons per laser pulse, and are logarithmically spaced to ensure a wide variety of input states within the dataset.
The digitizer is triggered on the signal from the photodiode and records 512 samples per trace, including 20 samples before the trigger.
In setting the attenuation, the external quantum efficiency is accounted for by measuring the efficiency of the complete system using quantum detector tomography \cite{Baalbergen2025_2}, such that $\mu=\eta\left<n\right>$ refers to the mean number of photons absorbed by the detector.

The digitizer triggering process introduces jitter due to its finite sample rate (\qty{5}{\giga\samples\per\second}), causing the trigger point to shift up to \qty{200}{\pico\second} compared to the true arrival time~\cite{teledyneManual}.
Because the jitter of the SNSPD system is much smaller than this offset~\cite{Sidorova2025, Schapeler2024, Calandri2016}, we correct for this digitizer-induced jitter by fitting a parabola to the top of the synchronization pulse to find the exact arrival time.
This arrival time is then used to shift the SNSPD signal in post-processing, by using the Fourier shifting theorem~\cite{weissteinFourierTransform}.

\section{Results}

\subsection{PCA: Optimal photon number resolution and interpretation}
\begin{figure*}
\centering
\includegraphics{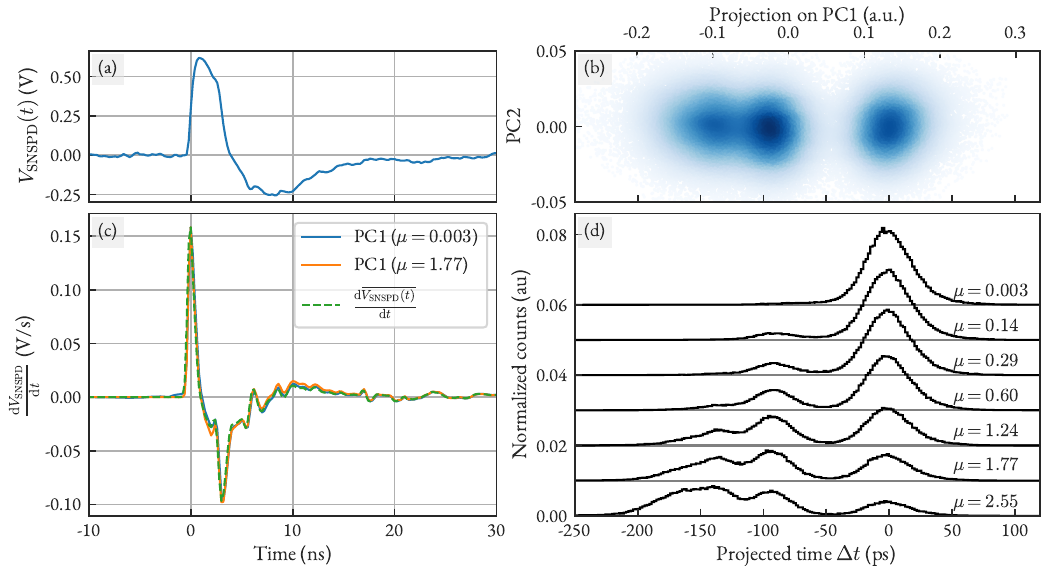}
\caption{Principal component analysis of detector pulses.
\textbf{(a)} Time trace of the output signal of the SNSPD.
\textbf{(b)} Example of the projection of the dataset at $\mu=1.77$ to the first two principal components.
\textbf{(c)} Comparison of the first principal component for $\mu=0.003$ (blue), $\mu=1.77$ (orange) and the average of the derivative of the output at $\mu=0.003$ (green dashed line).
\textbf{(d)} Projection of datasets of various mean photon numbers to the mean derivative of the $\mu=0.003$ dataset.
The results for projecting the data onto the first principal component of the $\mu=0.003$ dataset are indistinguishable from the projection on the average derivative.
}
\label{fig:results}
\end{figure*}

An example of a measurement trace is shown in \autoref{fig:results}a.
Fitting the PCA model yields 512 principal components, ranked according to the amount of variance in the dataset they capture.
The time traces can be projected onto the principal components to reduce the dimensionality of the data.
\autoref{fig:results}b shows an example of this projection onto the first two principal components for traces with a mean absorbed photon number $\mu=1.77$.
We find that the first principal component contains most relevant photon number information, which is apparent by the near-exclusive horizontal separation of clusters in \autoref{fig:results}b and this component's high explained variance (Appendix~\ref{sec:app:screeplot}).

The PCA is done separately for the different mean absorbed photon numbers $\mu$, and we find consistent results across all datasets.
The first principal components for the different datasets are all very similar to each other, as illustrated by the examples in \autoref{fig:results}c.
Also shown in the figure is the derivative of the mean trace of the dataset, which shows a remarkably strong resemblance to the first principal component.
This means that we can assign a photon number to an SNSPD pulse by calculating the projection of the trace to the mean derivative of the dataset.
To assign physical meaning to the values resulting from this projection, we approximate the detection traces for $n$ photons, $V_n(t)$, as time-shifted variants of the 1-photon trace, $V_1(t)$.
This approximation is consistent with the observation that higher photon-number events produce steeper rising edges, which to first order can be approximated as a small time-shift of an otherwise similar pulse shape.
Under this assumption, the projection $P$ can be written as
\begin{align}
P(V_n(t))=\int_\mathbb{R}\frac{\mathrm{d}\overline{V_1(t)}}{\mathrm{d}t}V_n(t)\mathrm{d}t \approx C\Delta t,
\label{eq:projection}
\end{align}
where $C$ is a constant and $\Delta t$ is a photon-number-dependent time shift.
For more information on this approximation, see Appendix~\ref{sec:app:timeshift}.
\autoref{eq:projection} shows that the approximated time shift of the projected trace can be calculated as $\Delta t = \frac{1}{C} P(V_n(t))$.
This projection to the mean derivative can thus be linked to a shift in time of the pulse under linear approximation, and yields the same results as PCA.
The results of the projection are shown in \autoref{fig:results}d.

The mean derivative varies depending on the photon-number distribution and therefore changes slightly for the different mean photon numbers $\mu$.
As the mean derivative is a linear combination of the derivatives of the different photon numbers, the exact choice of which derivative to use for the projection only introduces an offset in the calculated time shift.
In our analysis, we choose to use the mean derivative of the dataset corresponding to a mean photon number of $\mu = 0.003$.
In this set, 99.85\% of the light events are caused by the absorption of a single photon, making its mean trace highly similar to the mean trace of single-photon events, $\overline{V_1}$.
Projection onto the derivative of the mean 1-photon trace has the benefit that 1-photon traces will be projected to the origin, because for all differentiable functions the function and its derivative are orthogonal.
\autoref{fig:results}d shows the photon-number dependence for increasing mean photon numbers by the changing number and prominence of peaks, which agrees with the photon-number distribution of a coherent state as produced by a laser.

\subsection{Photon number resolvability}\label{sec:PNR}
To quantify the photon-number-resolving capability of a detector, we introduce a new confidence metric which depends on the overlap of consecutive photon-number peaks.
Although different performance metrics were used in previous literature~\cite{Schapeler2024, Schapeler2025}, we argue that they lack physical intuition, which we try to resolve with this new metric.
Our metric is based on the Bhattacharyya coefficient~\cite{Bhattacharyya1946}, which gives values between 0 and 1 depending on the amount of overlap between two underlying probability density functions.
We define the probability density function for an $n$-photon event as $G_{\left|n\right>}(t)$.
The confidence of resolving $n$ photons then scales with the amount of overlap between the corresponding $n$-photon peak, $G_{\left|n\right>}$, and the consecutive peak, $G_{\left|n+1\right>}$.
This consecutive peak $G_{\left|n+1\right>}$ is the limiting factor for the resolvability of $n$ photons, as it has the highest overlap with $G_{\left|n\right>}$~\cite{Sidorova2025}.
This results in the following confidence metric:
\begin{align}
C_{\left|n\right>\rightarrow \left|n+1\right>}&=1-\int_\mathbb{R}\sqrt{G_{\left|n\right>}(t)G_{\left|n+1\right>}(t)}\mathrm{d}t\label{eq:Cmetric}
\end{align}

The histograms in \autoref{fig:results}d show peaks due to 1-, 2- and more than 3-photon events.
To separate these peaks and quantify their overlap, a probability density function for the click probability as a function of projected time is needed.
We identify three contributions in the histogram:
\begin{align}
G_\textrm{total}(t, \mu) = P_1(\mu) G_{\left|1\right>}(t) + P_2(\mu) G_{\left|2\right>}(t) + P_{3+}(\mu) G_{\left|3+\right>}
\end{align}
Here $P_n(\mu)$ are the relative sizes of the different peaks in the histogram, the values of which are expected to follow the zero-truncated Poisson distribution.
The probability density functions $G_{\left|n\right>}(t)$ are well described by exponentially-modified Gaussians (EMGs)~\cite{Sidorova2017, schapeler2025_2, Sidorova2025}, and are assumed to be independent of the mean photon number $\mu$, which means that the shape of the EMG has to be fitted only once.

\begin{figure}
\centering
\includegraphics{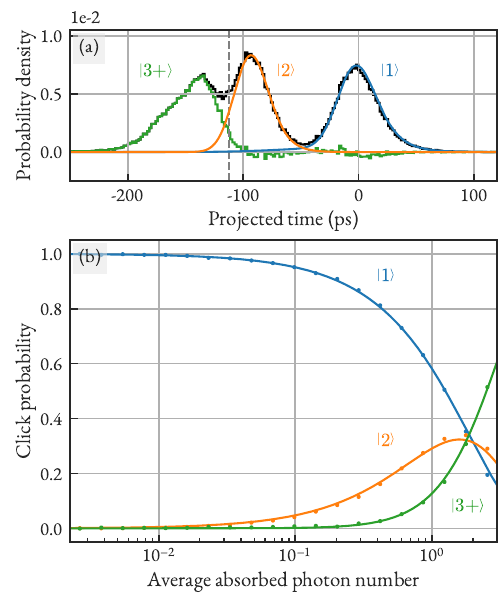}
\caption{Count statistics as a function of photon number.
\textbf{(a)} Measured histogram (black) and fitted exponentially modified gaussian distribution (blue and orange) for $\mu=1.77$.
The residuals are plotted in green. \textbf{(b)} Relative probability for the different photon numbers.
The probabilities for 1 and 2 photons are directly derived from the Gaussian fit.
The 3+ photon contribution is obtained by normalizing the residuals.
The solid lines give the expected zero-truncated Poisson distribution.
}
\label{fig:confidence}
\end{figure}

In order to find $G_{\left|1\right>}(t)$, we fit the $\mu=0.003$ data to a sum of an EMG and a Gaussian, to represent the 1-photon peak and the small amount of residual counts not contained in the 1-photon peak, respectively (see \autoref{fig:histogram_g1}).
For $\mu=0.003$ almost all observed events ($>99.8\%$) are caused by 1-photon events.
This fitted function $G_{\left|1\right>}(t)$ is used to fit amplitude of the 1-photon peak in the distributions for other values of $\mu$.
To fit the distributions for higher values of $\mu$, $P_1$ is first fitted after which the 1-photon peak is subtracted from the histogram and an EMG is fitted to the 2-photon peak.
The 3-or-more photon contribution is obtained by calculating the residual of this fit, and is normalized and labeled $G_{\left|3+\right>}$.
This residual cannot be fitted in an unambiguous way.
An example of the fit to the data at $\mu=1.77$ is shown in \autoref{fig:confidence}a.
The fitted values of $P_n(\mu)$ are used to calculate the detection probabilities for $n$ photons.
These detection probabilities are compared to the expected zero-truncated Poisson distribution in \autoref{fig:confidence}b, where a clear agreement can be seen.

By using the fitted probability density functions, we find the confidence for our measurements to be $C_{\left|1\right>\rightarrow \left|2\right>}=\num{0.81\pm0.01}$, and $C_{\left|2\right>\rightarrow \left|3+\right>}=\num{0.73\pm0.01}$.
The errors are found by random sampling from the fit parameters and calculating the variation in the confidence metric.
This confidence metric does not depend on the mean photon-number $\mu$, as the confidence is determined by using the fixed probability distributions.

\section{Discussion}
\begin{figure}
\centering
\includegraphics{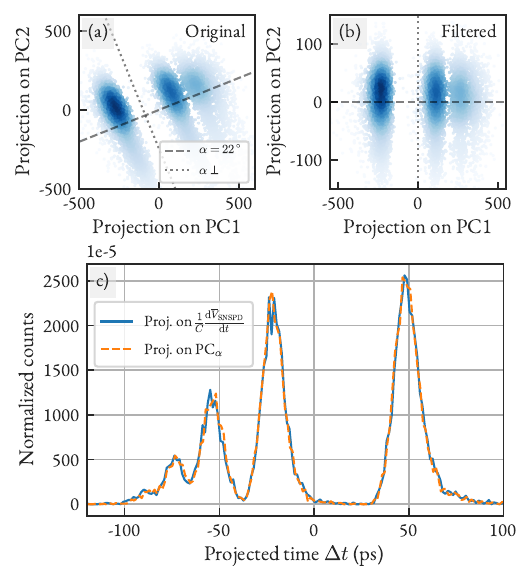}
\caption{Analysis of a subset of the dataset of Schapeler et al. corresponding to a mean photon number of $\mu=1.5$. \textbf{(a)} Projection on the first two principal components for the original data and \textbf{(b)} a filtered and downsampled version of the dataset.
The dashed and dotted lines indicate the projection and orthogonal axis used for assigning photon numbers to traces, respectively.
\textbf{(c)} Histogram of the data projected onto the average derivative (blue) and onto the effective principal component $PC_\alpha$ (orange, dashed). The latter is linearly scaled and shifted to align with the derivative projection, allowing direct comparison.
}
\label{fig:PCA_schapler}
\end{figure}

Previous methods to obtain photon-number information from time traces of SNSPDs used very high sample rates to digitize the electronic signal~\cite{Endo2021, Schapeler2024, Nicolich2019}.
We show that using our technique it is possible to obtain similar results using a digitizer with a modest sample rate of \qty{5}{\giga\samples\per\second}.
This result is not limited to our dataset.
We show this in \autoref{fig:PCA_schapler} by applying our method to the open-source dataset of Schapeler et al.~\cite{data-set-Schapeler}.
We use a subset of traces corresponding to a mean photon number of $\mu=1.5$ and create a digitally down-sampled variant by applying a first-order band-pass filter with cutoffs at 10 MHz and 2 GHz, and then downsampling from \qty{128}{\giga\samples\per\second} to \qty{4.92}{\giga\samples\per\second} by selecting every 26th sample point in each trace.
We find that after filtering, all photon number information shifts to the first principal component, ensuring the clusters are now horizontally separated instead of diagonally, as can be seen in \autoref{fig:PCA_schapler}a-b.
This shift of information can be explained by investigating the vector in the direction of the clusters, $PC_\alpha$, and the vector perpendicular to that, $PC_{\alpha\perp}$ (\autoref{fig:PCA_schapler}a).
The perpendicular vector $PC_{\alpha\perp}$ shows a constant offset, meaning that this direction captures the variance in the overall DC offset of the data, and no actual differences in the traces.
By applying a first-order band-pass filter, this DC offset gets removed from the data, collapsing all photon-number information to the first principal component.

To compare the photon-number resolving capability of the filtered dataset with that of the original dataset, we plot the projection histograms in \autoref{fig:PCA_schapler}c.
The original, unaltered data is projected onto the effective principal component $PC_\alpha$ defined by Schapeler et al, where the projection angle $\alpha$ is fitted for the specific subset of the data used here.
The filtered data is projected onto the mean derivative of the filtered dataset.
\autoref{fig:PCA_schapler}c shows that the photon-number resolving capability is equivalent for the original dataset and the digitally down-sampled variant, indicating that the significantly higher sample rate mainly enables better timing resolution in the signal trigger.

Moreover, we find that the effective principal component needed for the projection to extract photon-number information is again equivalent to the derivative of the mean trace.
\autoref{fig:PCA_schapler}c shows that also for this dataset, the photon number could be estimated by projecting the data to the mean derivative.
This broad applicability of our method across different setups can be explained by approximating the photon-number dependence as a shift in time of an unclassified $n$-photon trace relative to the mean trace.
An important distinction from similar methods is that our approach uses all available data points instead of one or two reference points, ensuring maximum information use.
The approximation of the photon-number dependence as a time-shift is justified as higher photon-number events result in steeper rising edges, which in first order can be approximated as a time-shift.
This result is not specific to our detector, as Sidorova et al.~\cite{Sidorova2025, Nicolich2019} theoretically predict steeper rising edges for higher photon-number absorption events in long nanowires with multiple hotspots, suggesting that our method is applicable to any such detector.

\begin{table*}
\centering
\ra{1.3}
\begin{tabular}{@{}lllllll@{}}
\toprule
Confidence metric     & \phantom{abc} & \multicolumn{2}{c}{This work (equation \ref{eq:Cmetric})} & \phantom{abc} & \multicolumn{2}{c}{Humphreys metric~\cite{Humphreys2015}}\\
\cmidrule{3-4}\cmidrule{6-7}
&& This work            & Schapeler et al.~\cite{Schapeler2024}&& This work & Schapeler et al.~\cite{Schapeler2024} \\ \midrule
$C_{1\rightarrow 2}$  && \num{0.83\pm0.01}   & \num{0.99\pm0.01} && \num{0.959\pm0.003}   & \num{0.999\pm0.002}\\
$C_{2\rightarrow 3}$  && \num{0.68\pm0.01}   & \num{0.99\pm0.01} && \num{0.881\pm0.003}   & \num{0.996\pm0.003}\\
$C_{3\rightarrow 4}$  && -                   & \num{0.85\pm0.02} && -                 & \num{0.96\pm0.01}\\
$C_{4\rightarrow 5+}$ && -                   & \num{0.7\pm0.2}   && -                 & \num{0.86\pm0.07}\\
\bottomrule
\end{tabular}
\caption{Different confidence metrics for the data of Schapeler at $\mu\approx1.5$ and data obtained in this work at $\mu\approx1.24$.
The left two columns show the confidence metric introduced in this work, \autoref{eq:Cmetric}, compared to the metric as proposed by Humphreys et al.~\cite{Humphreys2015} in the right two columns.}
\label{tab:confidence}
\end{table*}

To assess the photon-number resolving capabilities of different detectors, we introduced a confidence metric $C$ in \autoref{sec:PNR}.
This confidence metric enables direct comparison between our data and that of Schapeler et al., with the resulting confidences listed in \autoref{tab:confidence}.
We find that our detectors have a comparatively lower confidence $C_{1\rightarrow2}$ compared to peaks with similar separation in the histogram from Schapeler et al., this is due to a long tail in the 1-photon peak in our detector, causing a large overlap (see Appendix~\ref{sec:app:1photon_peak}).
The obtained confidences for the data of Schapeler et al. are the same for the $PC_{\alpha}$ projection of the original data and the mean derivative projection of the filtered data, demonstrating the effectiveness of our approach.
Our proposed metric can be considered setup-independent and provides a new standard for evaluating and comparing SNSPDs in terms of their photon-number resolving capabilities.

Other confidence metrics were proposed in previous works.
Our proposed metric is comparable to the metric as introduced by Humphreys et al.~\cite{Humphreys2015} for continuous output detectors.
We argue that the metric that we propose has advantages compared to the metric proposed by Humphreys et al.
The confidence metric proposed by Humphreys et al. describes the conditional probabilities for finding a photon number output, given an input state.
Therefore, this metric is not independent on the input state measured by the detector.
In order to characterize the behavior and capabilities of a detector, the used metrics should be independent of the details of the input state.
Our confidence metric measures how different the output signals are for different photon numbers, and thus does not depend on the input state.
A different method to determine the photon number resolving capabilities is proposed by Schapeler et al.~\cite{Schapeler2025}.
This method compares the separation of the click histograms to the width of these histograms and gives a single criterion for photon number resolvability.
The metric we propose gives a more detailed view on the specific capabilities of a detector and allows for comparing different detectors in more detail.

The confidence metric depends on the width of the photon-number peaks in the projection histogram, which is directly related to the jitter in the system, constraining the confidence.
The mean full-width at half-maximum of the 1-photon peak across all measured laser intensities for our system is $\Delta t_{\left|1\right>} \approx \qty{45(1)}{\pico\second}$, as measured from the timing histograms from \autoref{fig:results}d.
The laser pulse length (\qty{23}{\pico\second}) and the system jitter as measured by the manufacturer (\qty{35}{\pico\second}) together contribute $\Delta t_{\textrm{system}} \approx \qty{42(1)}{\pico\second}$ to the total measured jitter.
We attribute the small difference to jitter introduced by the photodiode and the amplifiers connected to the photodiode.
The SNSPD system is the main contributor to the total system jitter and thus the primary factor limiting the confidence.
It is therefore expected that reducing the jitter in the SNSPD system will increase the photon-number resolution according to our metric.

SNSPDs with intrinsic timing jitter as low as \qty{3}{\pico\second} were demonstrated~\cite{Korzh2020}.
However, these low jitter values are obtained in short-nanowire detectors, where only a single hotspot can form~\cite{Yang2007}.
For efficient detectors capable of photon number resolution, longer meandering wires are needed and the jitter will mainly be determined by geometric jitter~\cite{Calandri2016}.
This means that there is an intrinsic trade-off between high detection efficiencies and the best photon number resolution obtainable.

The mean derivative projection method introduced to obtain optimum photon-number resolution can be implemented in an FPGA.
This can be performed by using a device similar to the digitizer used in this work.
Our digitizer (Teledyne ADQ7), is capable of $300\times10^6$ pulse samples per second using pulse detection firmware.
Using custom-made peak analysis, our device could be setup to perform our analysis to obtain the photon number\cite{teledyneFWPD, teledyneArtOfPeak}.
Previous work on computationally similar calculations show that even for relatively low speed FPGAs the latency can be on the order of \qty{10}{\micro\second}~\cite{Pang2013}.
The first step of our original technique corrects for the digitizer-induced jitter by using a curve-fitting algorithm to find the trace-specific jitter, and corrects for this by Fourier shifting the trace.
Although effective, this step is computationally expensive.
We introduce an alternative method to correct for the digitizer-induced jitter and compute the projected time $\Delta t$ in a way that requires minimal computations (Details of this method are given in Appendix~\ref{sec:app:optimized_projection_method}).
This technique paves the way for real-time photon-number classification when implemented in an FPGA and enables feedforward in quantum devices as a function of photon number~\cite{mario2017, Lopes2010, Mittal2020}.

\section{Conclusions}
We have demonstrated a systematic framework for identifying and quantifying photon-number resolution in SNSPDs using principal component analysis.
Our results show that the first principal component corresponds to the derivative of the mean response trace, allowing a simple and physically interpretable way to extract the photon-number information from digitized time traces.
This method provides a direct and scalable route to photon-number determination that can be implemented with modest hardware requirements.
To benchmark the photon-number-resolving performance, we introduced a confidence metric based on the overlap of consecutive photon-number peaks.
The confidence metric can be compared to a fidelity and provides an intuitive, quantitative measure of resolvability.
The results demonstrate that in our setup the confidence is primarily limited by the SNSPD system jitter, guiding the potential for improvement of the photon-number resolving capability.
Finally, we show that this method generalizes across datasets and experimental setups, and can be implemented efficiently on FPGA hardware, paving the way for real-time photon-number classification and feed-forward operation in quantum photonic systems.

\begin{acknowledgments}
The authors thank Mio Poortvliet and Wolfgang Löffler for use of the detector system.
The authors thank Single Quantum B.V. for proving a pre-production engineering sample detector.
This research was made possible by financial support from the Dutch Research Council (NWO) under Project 19716.
I.K. and E.vN. acknowledge support from the Dutch National Growth Fund (NGF), as part of the Quantum Delta NL programme.
\end{acknowledgments}

\appendix
\section{Scree plot of principal component analysis}\label{sec:app:screeplot}
\begin{figure}
\centering
\includegraphics{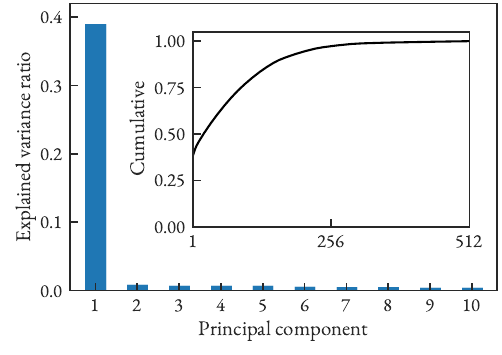}
\caption{
Scree plot showing the explained variance of the first 10 principal components in the principal component analysis for $\mu=1.77$. The inset shows the cumulative explained variance as a function of the number of principal components.
}
\label{fig:screeplot}
\end{figure}

\autoref{fig:screeplot} shows the scree plot for our principal component analysis, where the principal components are naturally ordered by their explained variance.
The first component accounts for most of the variance and is clearly more significant than the subsequent ones, indicating that most of the photon-number information is contained in PC1.
This is corroborated by the inset, which shows the cumulative explained variance, indicating that about 200 principal components are needed to capture the remaining noise in the experimental data.

\section{Time shift approximation}\label{sec:app:timeshift}
\begin{figure}
\centering
\includegraphics{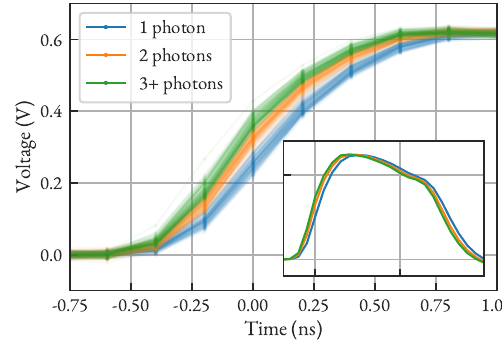}
\caption{Difference in rising edge of SNSPD traces.
Measured traces, sampled at \qty{5}{\giga\samples\per\second}, are colored according to the classification as 1 photon (blue), 2 photons (orange) and 3 or more photons (green). The inset shows mean traces for a broader range, indicating that the time shift persists over a larger range.
}
\label{fig:rising_edge}
\end{figure}

The classification of pulses to different photon numbers is done by projection of the detection trace to the mean derivative of all SNSPD traces.
This projection can be linked to a time-shift $\Delta t$ if a linear approximation is used.
For any differentiable function $f:\mathbb{R}\rightarrow\mathbb{R}$, the function can be approximated to first order in $\Delta t$ as
\begin{align}
f(t+\Delta t)&\approx f(t)+\Delta t \frac{\mathrm{d}f(t)}{\mathrm{d}t}.
\end{align}
Assuming that the detection trace for $n$ photons can be approximated as a time-shifted version of the trace for 1 photon (see \autoref{fig:rising_edge}), we write
\begin{align}
V_n(t)&=V_1(t+\Delta t)\approx \overline{V_1(t)}+\Delta t \frac{\mathrm{d}\overline{V_1(t)}}{\mathrm{d}t}.
\end{align}
Where $V_n(t)$ is a detection trace for $n$ photons and $\Delta t$ is the time shift of $V_n$ compared to $V_1$.
Due to experimental noise, $V_1(t)$ is best estimated by averaging over all 1-photon traces, denoted by $\overline{V_1(t)}$.
Projecting $V_n(t)$ onto $\frac{\mathrm{d}\overline{V_1(t)}}{\mathrm{d}t}$, we find
\begin{align}
P(V_n(t))&=\int_\mathbb{R}\frac{\mathrm{d}\overline{V_1(t)}}{\mathrm{d}t}V_n(t)\mathrm{d}t\notag\\
&\approx\int_\mathbb{R}\frac{\mathrm{d}\overline{V_1(t)}}{\mathrm{d}t}\left(\overline{V_1(t)}+\Delta t \frac{\mathrm{d}\overline{V_1(t)}}{\mathrm{d}t}\right)\mathrm{d}t.
\intertext{Since for all differential functions the function and its derivative are orthogonal, we can write the projection as}
P(V_n(t))&\approx\Delta t\int_\mathbb{R}\left({\frac{\mathrm{d}\overline{V_1(t)}}{\mathrm{d}t}}\right)^2\mathrm{d}t\notag\\
&=C\Delta t,
\end{align}
where $C$ is a normalization constant.

It is important to note that this first order approximation implies that projection to the time derivative is equivalent to measuring a difference in arrival time, if the shape of the traces are similar enough.
Here, we use $\overline{V_1}$ as a reference trace, but under the assumption that the different photon-number traces can be approximated as time-shifted versions of each other, this reference could also be the mean trace of the entire dataset.

\section{Timing histogram for the 1-photon peak}\label{sec:app:1photon_peak}
\begin{figure}
\centering
\includegraphics{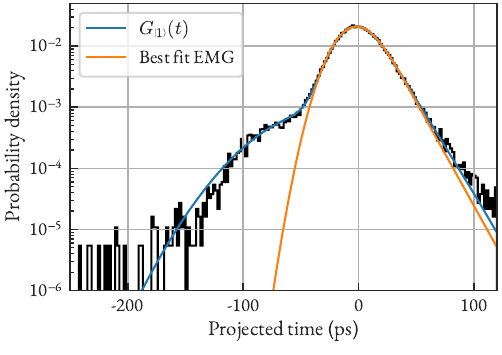}
\caption{Click histogram for an average photon number of $\mu=0.003$.
The blue line shows the fitted function $G_{\left|1\right>}(t)$ consisting of an EMG and an extra gaussian.
The orange line shows the best fitted EMG function.
The combined fit clearly shows better agreement with the measured data capturing also the shoulder near \qty{-100}{\pico\second}.
}
\label{fig:histogram_g1}
\end{figure}

The 1-photon peak of our detector consists of a main peak described by an EMG and a broad shoulder at the negative projected time.
For an average absorbed photon number of $\mu=0.003$, more then 99.8\% of events should be 1-photon events.
In order to make a good estimate of the true 1-photon response, we also fitted the shoulder of the count histogram.
\autoref{fig:histogram_g1} shows the timing histogram for this photon number in a semi-log scale with both the best EMG function (orange) and our best found function $G_{\left|1\right>}(t)$ (blue).
A clear improvement in the description of the 1-photon peak can be seen for the EMG + gaussian fit compared to the best fitted EMG.

\section{Hybrid method for SNSPD trace projection and digitizer jitter correction}
\label{sec:app:optimized_projection_method}
To decrease the computational load of our trace projection method and allow efficient implementation in an FPGA, we introduce an alternative method to correct for digitizer-induced jitter that uses unprocessed, non time-shifted traces.
We start by determining the trace-specific digitizer-induced jitter using the time shift approximation introduced in Appendix~\ref{sec:app:timeshift} by calculating the time shift of each measured synchronization signal with respect to the mean synchronization signal.
Then, after calculating the time shift of each measured SNSPD signal with respect to the mean 1-photon trace, we subtract the corresponding time shift of the synchronization signal to recover the projected time. Mathematically, we project a measured pair of SNSPD and synchronization signal,
\begin{align}
\vec{v}_\text{measurement}=\left(\vec{V}_\text{SNSPD},\vec{V}_\text{PD}\right)^T,
\end{align}
onto
\begin{align}
\vec{v}_\text{projection}=\left(\frac{1}{C_{\left|1\right>}}\partial\overline{V_{\left|1\right>}},-\frac{1}{C_\text{PD}}\partial\overline{V_{\text{PD}}}\right)^T,
\end{align}
to find the projected time
\begin{align}
\Delta t = \vec{v}_\text{measurement}\cdot\vec{v}_\text{projection}.
\end{align}
Here, $\partial\overline{V_{\left|1\right>}}$($\partial\overline{V_\text{PD}}$) is the mean derivative of the SNSPD (photodiode), $C_{\left|1\right>}$($C_\text{PD}$) is a normalization constant calculated in Appendix~\ref{sec:app:timeshift} for the SNSPD (photodiode) trace.
The projection vector is a constant for each measurement system and can be calculated before running the SNSPD.

\section{Comparison between PCA and average derivative method for a research detector}
To show the wide applicability of our method, we measured time traces from a fiber coupled early-stage developmental prototype detector from Single Quantum with an approximate $60\%$ external and $100\%$ internal quantum efficiency\footnote{The efficiency of the early-stage developmental prototype provided by Single Quantum has not been optimized for system detection efficiency at the experimental wavelength.}.
This detector is mounted in a closed cycle cryostat at \qty{4.3}{\kelvin} and connected to a cryogenic amplifier (CMT-BA1) at \qty{59}{\kelvin} via a \qty{5}{\dB} attenuator (minicircuits BW-S5W2+).
The signal from the cryogenic amplifier is further amplified using a room temperature amplifier (Minicircuits ZFL-1000LN+) and sampled using the same method as described in the main text.
This complete system contains low frequency noise causing the baseline of the detector output to vary for all pulses.
Furthermore, the detector is biased at a relatively low current due to the high temperature of the system, limiting the signal-to-noise ratio of the detector output.

\begin{figure}
\centering
\includegraphics{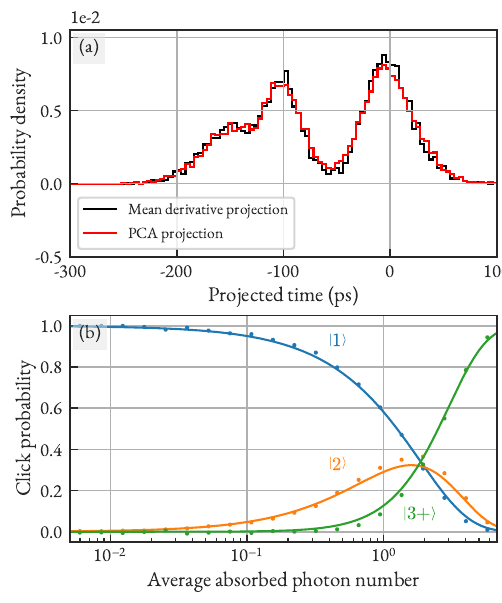}
\caption{
Comparison of count statistics as determined by PCA and mean derivative projection for the Single Quantum detector.
\textbf{(a)} Measured mean derivative projection histogram (black) and PCA projection histogram (red) for $\mu=1.36$.
\textbf{(b)} Relative click probability for the different photon numbers.
The probabilities for 1 and 2 photons are directly derived from EMG fits.
The solid lines give the expected zero-truncated Poisson distribution.
}
\label{fig:confidence_SQ}
\end{figure}

To be able to perform PCA, the DC component of each trace is removed.
This is done by subtracting the mean value from the first 50 data points of the data trace.
After removing this low frequency noise in the system, the projection on the first two principal components show a similar trend as seen in the paper from Schapeler et al.~\cite{Schapeler2024}.
The optimal projection angle is fitted and the resulting projection histogram for $\mu=1.36$ is shown in \autoref{fig:confidence_SQ}a (red).
From the original raw traces, without doing any pre-processing, the average derivative is calculated and the projection on each trace is calculated.
The mean derivative projection for for $\mu=1.36$ is also plotted in \autoref{fig:confidence_SQ}a (black), allowing a direct comparison of the measured timing histograms.
This Figure shows that the mean derivative projection method results in more prominent peaks.
Additionally, the mean derivative projection method requires far less processing steps to achieve this result.
Even with a noisy dataset, optimal photon number resolution is obtained using this method, as demonstrated by the click probabilities in \autoref{fig:confidence_SQ}b.
The click probabilities are obtained by fitting EMG curves to both the 1 and 2 photon peaks using the same method as in the main text.

%

\end{document}